\documentclass{aa}
\usepackage{graphicx}
\def\ssim{\setbox0=\hbox{$\propto$}%
\setbox1=\hbox{$<$}\dimen0=\ht1%
\advance\dimen0by-1.2pt\,\lower.6\dimen0%
\copy0\kern-\wd0\raise.4\dimen0\copy1 \,}

\def\gsim{\setbox0=\hbox{$\propto$}%
\setbox1=\hbox{$>$}\dimen0=\ht1%
\advance\dimen0by-1.2pt\,\lower.6\dimen0%
\copy0\kern-\wd0\raise.4\dimen0\copy1\,}

\def\lambdab{\lambda\mkern-9mu\lower1.2pt\hbox{$\mathchar'26$}}%

\begin{document}
   \title{Research Note:\\Rotation and the wind
 momentum--luminosity relation for extragalactic distances}

 \author{A. Maeder}

     \institute{Geneva Observatory CH-1290 Sauverny, Switzerland
              email: Andre.Maeder@obs.unige.ch
                      }

   \date{Received ...; accepted ...}

\abstract{ The effects of axial stellar rotation
on the wind--momentum relation (WLR) for determining
the extragalactic distances are investigated. 
Despite the fact that the mass loss rates grow
quite a lot with rotation, remarkably
 the effects on the WLR are  found to be very small on the average.
As an example, for an  average orientation angle
between  the rotation axis and the line of sight,
the luminosity would be overestimated
by  5.9 \%   for   a star rotating at 90\% of its
break--up rotational velocity. Different orientation
angles between the rotation axis and the line of sight
produce some limited scatter.
               }

   \maketitle
%

\section{Introduction}
There is an important  relation  between the mechanical 
wind momentum $\dot{M} v_{\infty}$ of  the mass outflow
and the luminosity $L$ of hot stars predicted
by the theory of radiative winds (e.g.  Kudritzki et al.
\cite{kudr89}; Kudritzki \cite{kudr00}; Kudritzki \& Puls
\cite{kudrpuls00}). This relation writes 

\begin{equation}
\dot{M} v_{\infty} \; \propto  \; R^{-0.5} L^{\frac{1}{\alpha}} \; .
\end{equation} 

\noindent
where $\dot{M}$ is the mass loss rate, 
 $v_{\infty}$ the terminal velocity and $R$ the stellar radius
The parameter $\alpha$ is equal to about $\frac{2}{3}$, it
is a so--called force--multiplier representing the power law 
exponent  of the distribution of line strengths of the 
spectral lines driving the wind. The observations of
O--type stars in the Galaxy and the LMC have shown
the existence of a tight relationship 
between the wind
momentum and the luminosity (Puls et al. \cite{puls96}). 
More recently, the existence of a wind momentum--luminosity
 relation (WLR) 
has also been confirmed  for A-- and B--type supergiants
(Kudritzki et al. \cite{kudr99}). 

The potentialities of the WLR for  the determination of
extragalactic distances are claimed to be great 
(Kudritzki \cite{kudr00}). The WLR provides
absolute luminosities from
direct spectroscopic measurements of
the  mass loss rates and of the terminal velocities obtained from
the line profiles. Kudrizki (\cite{kudr00}) states
that this  method ``may allow independent distance moduli 
to be obtained with an accuracy of 10\% out to the Virgo and Fornax 
clusters of galaxies''.

Rotation has been shown  to increase
significantly the mass loss rates of a star of a given mass and
luminosity, i.e. up to a factor 2 and 4 for stars of 40 and 60 M$_{\odot}$,
(Maeder \cite{MIV}; Maeder and Meynet \cite{MMVI}).
For stars close to the Eddington--limit, rotation may even  
increase the mass loss rates by orders of magnitude with respect
to the non rotating case. Also, rotation modifies the terminal
velocities  $v_{\infty}$ and the wind densities 
according to the colatitude $\vartheta$.
In view of these large effects of rotation, we may wonder 
about the effects of rotation
on the WLR. This may also be particularly useful if the proportion
of fast rotators is higher in lower metallicity galaxies, as
suggested by some recent observations (Maeder et al. \cite{maegremer};
Keller \cite{kell00}). 

\section{Rotational effects on the WLR}

\subsection{Classical case} 

In the classical case without rotation, the mass loss rates and the
terminal velocities behave as (Puls et al. \cite{puls96}; Kudritzki
\cite{kudr00})

\begin{eqnarray}
\dot{M} \; \propto \; L^{\frac{1}{\alpha}} [M(1-\Gamma)]^{1-\frac{1}{\alpha}}
\; ,\\
v_{\infty} \;\propto \; 2.24 \frac{\alpha}{1-\alpha} 
\left[\frac{GM(1-\Gamma)}{R} \right]^{0.5} \; .
\end{eqnarray}

\noindent
$\Gamma = 
\frac{\kappa_{\mathrm{es}} L}{4 \pi c GM}$  is the Eddington factor
 with $\kappa_{\mathrm{es}} $ the electron--scattering
opacity. The WLR is thus 

\begin{equation}
\dot{M} v_{\infty} \; \propto \;  \frac{1}{R^{0.5}} L^{\frac{1}{\alpha}} [M(1-\Gamma)]^{\frac{3}{2}-\frac{1}{\alpha}} \; .
\end{equation}

\noindent
Thus, with the value $\alpha = 2/3$ typical for  O--stars, one 
obtains the  WLR relation given by Eq. (1).

\subsection{Local WLR}

In the case of rotating stars, the stellar shape  defined
by the Roche model has to be accounted
for. For a star of angular velocity
$\Omega$, the surface equation is a function of a parameter 
$\omega^{2} = \frac{\Omega^{2} R^{3}_{\rm{eb}}}{GM}$,
 where $\Omega$ is the angular velocity
and $\omega$ the  fraction of the angular velocity
at break--up. R$_{\mathrm{eb}}$ is the equatorial radius at break--up.
The von Zeipel theorem  is used for defining the local flux,
with possibly some corrections for differential rotation
(Maeder \cite{MIV}). 
The Eddington factor $\Gamma_{\Omega}(\vartheta)$ 
in a rotating star must be defined
as the ratio of the \textit{local} flux to the \textit{local}
limiting flux
(Maeder \& Meynet \cite{MMVI}). We recall that
in the framework of the wind theory, the  Eddington factor 
appearing in the various expressions 
is that for electron scattering 
opacity   $\kappa_{\mathrm{es}} $,
 since the true opacities are expressed
as a function of  $\kappa_{\mathrm{es}} $  by means of the 
force multipliers (Kudritzki \& Puls
\cite{kudrpuls00}). In this specific context,
 $\Gamma_{\Omega}$ is the same at all latitudes

\begin{equation} 
\Gamma_{\Omega} = \frac{\Gamma}{1 - \frac{\Omega^2}{ 2 \pi G 
\rho_\mathrm{m}}} \; ,
\end{equation}

\noindent
where $\Gamma$ is the Eddington factor  defined above and
 $\rho_{\rm{m}}$ is the average density inside the surface
equipotential. We also stress that 
the critical velocity is not given by 
$v_{\rm{crit}} = \left[
\frac{GM}{R} (1-\Gamma) \right]^\frac{1}{2}$ 
as often considered, which applies to uniformly bright stars.
In a rotating star with a variable local 
T$_{\mathrm{eff}}$ in latitude, more care is needed.
 Due to the von Zeipel theorem,
the critical velocity is equal to $v_\mathrm{crit} =  
(\frac{GM}{R_{\mathrm{eb}}})^ {\frac{1}{2}}$ as long as $\Gamma 
\leq 0.639$, while above it, the critical 
velocity progressively decreases to zero 
(Maeder \& Meynet \cite{MMVI}).

 In a rotating star, the mass flux
is not constant over the stellar surface.
 At a colatitude $\vartheta$ the mass flux,
i.e. the amount of mass loss $\Delta\dot{M}$ by surface element 
$\Delta \sigma$ is (Maeder \& Meynet \cite{MMVI})

\begin{eqnarray}
\frac{\Delta\dot{M}(\vartheta)}{\Delta \sigma}
\propto  A \; \left[\frac{L(P)}{4\pi
GM_\star (P)}\right]^{\frac{1}{\alpha}}
\frac{g_\mathrm{eff} [1+\zeta(\vartheta]^{\frac{1}{\alpha}}}
{[1 - \Gamma_{\Omega}(\vartheta)]^{\frac{1}{\alpha}-1}} \nonumber \\[2mm] 
 \quad {\mathrm{with}} \quad
A = \left(k\alpha\right)^{\frac{1}{\alpha}} \left(
\frac{1-\alpha}{\alpha}\right)^{\frac{1-\alpha}{\alpha}} \; ,
\end{eqnarray}

\noindent
$L(P)$ and $M_{\star}(P)$ are the luminosity and effective mass
inside an equipotential (cf. Maeder \& Zahn \cite{MZ98}), here 
 the surface equipotential. One has
\begin{equation}
M_{\star}(P) = M \left( 1 - \frac{\Omega^2}
{2 \pi G \rho_{\rm{m}}}  \right) \; .
\end{equation}
\noindent
The effective gravity $\vec{g}_\mathrm{eff}$ includes the gravitational
and centrifugal acceleration, while the total gravity 
$\vec{g}_\mathrm{tot} =  \vec{g}_\mathrm{eff} +\vec{g}_\mathrm{rad}$
also accounts for the radiative acceleration.
If we neglect  the deviations from the von Zeipel theorem due 
to differential rotation,
the term $\zeta(\vartheta)$  in Eq. (6) is equal to zero. 

In a rotating star, the terminal
velocity $v_{\infty}(\vartheta)$ depends 
on the colatitude $\vartheta$, it behaves like
(Lamers \& Cassinelli \cite{LamCass99})

\begin{eqnarray}
v_{\infty}(\vartheta)  \; \propto \; [g_{tot}(\vartheta)
 R(\vartheta)]^{\frac{1}{2}} \nonumber \\[2mm]
\;  = \; \left[g_{\mathrm{eff}}(\vartheta)(1-\Gamma_{\Omega}) R(\vartheta)
\right]^{\frac{1}{2}} \; .
\end{eqnarray}

\noindent
For $g_{\mathrm{eff}}(\vartheta)$, we shall take below the
radial component. Thus, 
 we shall omit a factor $\cos\epsilon$
(cf. Maeder \cite{MIV}), where
$\epsilon$ is a small angle 
between the radial vector and the effective gravity.
At most, $\cos\epsilon$ deviates from unity by 3\%. We get
for the local mass flux--velocity--luminosity relation

\begin{eqnarray}
\frac{\Delta\dot{M}(\vartheta) v_{\infty}(\vartheta) }{\Delta \sigma}
\propto  \left[\frac{L(P)}{4\pi
GM_\star (P)}\right]^{\frac{1}{\alpha}}
\frac{g_\mathrm{eff}^{\frac{3}{2}}(\vartheta)
 \; R^{\frac{1}{2}}(\vartheta)}
{[1 - \Gamma_{\Omega}(\vartheta)]^{\frac{1}
{\alpha}- \frac{3}{2}  }}   \; .
\end{eqnarray}

\noindent
The modulus of the effective gravity in the radial direction is

\begin{eqnarray}
g_\mathrm{eff}(\vartheta) = 
\frac{GM}{R^2(\vartheta)}\left[1 - w^{2}(\vartheta) \right] \; ,
\end{eqnarray}

\noindent
with $w^{2}(\vartheta) = \frac{\Omega^{2} R^{3}
(\vartheta)}{GM} \sin^{2}\vartheta $. The local  WLR  becomes

\begin{eqnarray}
\dot{M}(\vartheta) \; v_{\infty}(\vartheta) \; \propto \; 
\frac{L^{\frac{1}{\alpha}}(P)}{R^{\frac{1}{2}}(\vartheta)} 
\nonumber  \\[2mm]
\frac{\left[ M(1-\Gamma_{\Omega}) 
\right]^{\frac{3}{2} -\frac{1}{\alpha}}
\left[ 1- w^{2}(\vartheta) \right]^{\frac{3}{2}}}
{\left( 1 - \frac{\Omega^2}
{2 \pi G \rho_{\rm{m}}}  \right) ^{\frac{1}{\alpha}}} \; .
\end{eqnarray}

\noindent
Expressing $1-\Gamma_{\Omega}$, we finally get for the  local  WLR

\begin{eqnarray}
\dot{M}(\vartheta) \; v_{\infty}(\vartheta) \; \propto \; \nonumber
\\[2mm]
\frac{L^{\frac{1}{\alpha}}(P)}{R^{\frac{1}{2}}(\vartheta)} 
\left[ M(1-  \frac{\Omega^2}{2 \pi G \rho_{\rm{m}}} 
- \Gamma)
\right]^{\frac{3}{2} -\frac{1}{\alpha}} f_{\omega}(\vartheta) 
\end{eqnarray}

\noindent
with  $f_{\omega}(\vartheta)$ given by 

\begin{equation}
 f_{\omega}(\vartheta)= \frac{
 \left( 1-w^{2}(\vartheta) \right)^{\frac{3}{2}}}
{\left( 1 - \frac{\Omega^2}
{2 \pi G \rho_{\rm{m}}}  \right) ^{\frac{3}{2}}} \; .
\end{equation}

\subsection{Average orientation}

The previous expression gives the local WLR for a surface
element of the star at a given colatitude 
$\vartheta$. We have to examine what  is  the average  WLR 
 and also how it depends on the
 orientation  angle $i$ between the rotation axis and the
line of sight.
 Let us firstly examine Eq. (13) for an average
orientation.  In the case of  a random distribution of the
orientation  angles  $i$, the average angle $<i>$
is equal to 57.3 degrees. Let us consider for simplicity
an average orientation  corresponding to the vertical
direction on a surface element at the zero of the
Legendre Polynomial
 $P_{2} (\cos (\vartheta)) = 0$, i.e. for
$\sin ^{2} (\vartheta) = 2/3$, which corresponds to a colatitude
of 54.7 degrees.  This colatitude also corresponds
to that of the average stellar radius in models
with rotation (Meynet \& Maeder \cite{MMI}).
 Let us call $R_{\mathrm{o}}$
this average stellar radius at the considered rotation,
$R_{\mathrm{o}}$ is as a matter of fact the best estimate 
of the average radius of a rotating star.
  We have at the numerator of Eq. (13)

\begin{eqnarray}
1-w^2 = 1 - \frac{2}{3}
\frac{ \Omega^2 R_{\mathrm{o}}^3}{GM} \nonumber \\[2mm]
= \; 1 - \frac{4}{9} \frac{\Omega^{2} 
R^{2}_{\rm{pb}}}{v^{2}_{\rm{crit}}}
\left( \frac{R_{\mathrm{o}}}{R_{\rm{pb}}} \right)^{3}
= \; 1  - \frac{4}{9} \left( \frac{v}{v_{\rm{crit}}}\right)^2 
\varphi (\omega) \; ,
\end{eqnarray}

\noindent
where  $\varphi(\omega)$ is given by

\begin{equation}
\varphi(\omega) = \left( \frac{R_{\rm{pb}}}{R_{\rm{e}}}
\right)^{2} 
\left (\frac{R_{\mathrm{o}}}{R_{\rm{pb}}} \right)^{3} \; .
\end{equation} 

\noindent 
$R_{\rm{pb}}$ is the polar
 radius at break--up, $R_{\rm{e}}$
is the equatorial radius at the considered rotational velocity 
$v=\Omega R_{\mathrm{e}}$.
The ratio $\varphi(\omega)$ is derived from the Roche model,
it is always slightly smaller than 1.0.
For the denominator of Eq. (13), we have
(Maeder and Meynet \cite{MMVI})

\begin{equation}
\frac{\Omega^2}{2 \pi G \rho_{\mathrm{m}}}=
\frac{4}{9} \frac{v^2}{v_\mathrm{crit}^2} V^{\prime}(\omega)
\frac{R^2_{\mathrm{pb}}}{R^2_{\mathrm{e}}} \; .
\end{equation}

\noindent
The volume $V^{\prime}(\omega)$ is the ratio of the actual volume
of the rotating star to the spherical volume of radius R$_{\mathrm{
pb}}$. The relation between $\frac{\Omega^2}{2 \pi G \rho_{\rm{m}}}$
and the ratio $(\frac{v}{v_\mathrm{crit}})^2 $  is almost linear 
(Maeder and Meynet, \cite{MMVI}) and to first order one has
 $\frac{\Omega^2}{2 \pi G \rho_{\rm{m}}}
= \frac{4}{9} (\frac{v}{v_\mathrm{crit}})^2$.
For the average orientation defined above,
 the wind momentum--luminosity relation is thus

\begin{eqnarray}
\dot{M}\; v_{\infty} \; \propto \; 
\frac{L^{\frac{1}{\alpha}}(P)}{R^{\frac{1}{2}}_{\mathrm{o}}} 
\left[ M(1-  \frac{\Omega^2}{2 \pi G \rho_{\rm{m}}} 
- \Gamma)
\right]^{\frac{3}{2} -\frac{1}{\alpha}} f_{\omega} 
\end{eqnarray}

\noindent
For OB stars, the force multiplier $\alpha$ is very close
to 2/3 and one is left with

\begin{eqnarray}
\dot{M}\; v_{\infty} \; \propto \; 
\frac{L^{\frac{1}{\alpha}}(P)}{R^{\frac{1}{2}}_{\mathrm{o}}} 
 f_{\omega} 
\end{eqnarray}

\noindent
Eqs. (14) and (16) enable us to calculate the fraction 
$f_{\omega}$ in Eq. (17).
The variations of  $f_{\omega}$  as a function 
of $\frac{v}{v_\mathrm{crit}}$ are shown in Fig. 1.
We see that, up to a fraction 0.6 of the critical velocity,
$f_{\omega}$ does not deviate from 1 by more than 1 \%,
up to a fraction 0.8 of the critical velocity the maximum deviation
is 4 \%, and for a fraction 0.9 it is 9 \%.
Since in Eq. (18), the luminosity appears with an exponent
$1/ \alpha$, the effects on the luminosity determinations
are even smaller. As an example, for  
a fraction 0.8 of the critical velocity, the luminosities
which would be derived from the usual Eq.(1) would be too large by
 2.6 \%, and by 5.9 \% for a fraction 0.9 of the critical velocity.
This means that  for  given $\dot{M} v_{\infty}$
and radii, the determinations of the luminosity 
is not very much modified by rotation, even very fast.
The function   $f_{\omega}$ behaves like
$f_{\omega} = 1.17  
\left( \frac{v}{v_{\mathrm{crit}}} \right) ^6 $
with deviations smaller than 0.002 up to $\frac{v}{v_{\mathrm{crit}}}
= 0.97$.

It is remarquable that, while all quantities in Eq. (18)
are substantially influenced by rotation (in particular the mass
loss rates),
 the overall relation
between the actual mass loss rates, terminal velocities, 
luminosities and radii
is keeping the same for OB stars.
We conclude that,  because of compensation effects,
\emph{ the effects of rotation are remarkably absent 
in the average wind--momentum relation for OB stars}.

\begin{figure}
\resizebox{\hsize}{!}{\includegraphics{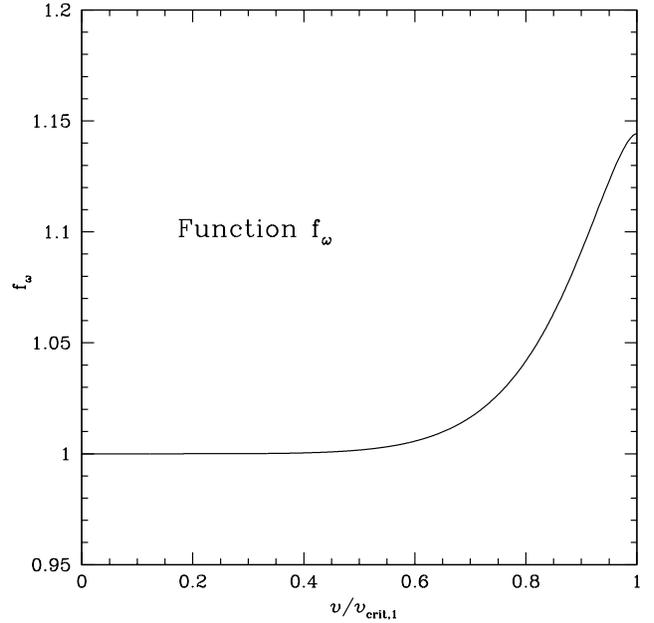}}
\caption{The function $f_{\omega}$ which expresses the
deviations from the usual WLR is shown in terms of the fraction
of the critical velocity, (cf. Eq. 18). }
\label{car00}
\end{figure}

\subsection{Orientation effects}

We need also to estimate  the size of the effects of orientation, which 
 could produce some
scatter around the mean WLR given above. Let us consider the extreme 
cases of regions at the pole and at  the equator.
Firstly, we notice that  at the pole, the values of 
$f_{\omega}(\vartheta)$ become bigger than 1.0 for growing
rotations. For example,  for  values of the fraction 
of the break--up angular velocity 
$ \omega = 0.5, 0.8$, one respectively  has
$f_{\omega}(\mathrm{pole}) = 1.086$ and  $1.305$. 
On the contrary, at the equator the values 
$f_{\omega}(\vartheta)$ becomes smaller than 1.0 for growing
rotations. For example, for 
 $ \omega = 0.5, 0.8$, one respectively  has
$f_{\omega}(\mathrm{equator}) = 0.952 $ and $0.796$. This means that 
locally, the deviations from
the WLR due to the factor $f_{\omega}(\vartheta)$ are not negligible.

Now, we must account that the observable effects are integrated
over the visible part of the stellar surface, this produces
an averaging  of regions with different values of
 $f_{\omega}(\vartheta)$.
 Let us do it for  the extreme cases of
 pole--on and of  equator--on observers.
 For an observer who is seeing pole--on a rotating star
 with $ \omega =$ 0.5, 0.8 and 0.9, the apparent 
surface weighted values
of $f_{\omega}$(pole--on) are 1.021, 1.068 
and 1.103 respectively.
This shows that the luminosities derived by the
application of the usual Eq. (1) are too large by
1.4, 4.5 and 6.8 \% respectively.

For an observer who is seeing  equator--on a rotating star
 with $ \omega =$ 0.5, 0.8 and 0.9, the apparent 
surface weighted values
of $f_{\omega}$(equator--on) are 0.987,
0.934 and 0.878 respectively. 
These last values imply that the luminosities which
 would be derived by the application of the usual
 Eq. (1) would be too small by 0.9, 4.5 and 8.3 \%
respectively.
This shows that, even for very fast rotations,
the deviations from the average WLR relation, due to
orientation effects, are  relatively small.

Finally, we may also consider a weighting according 
to the local flux emitted. The results are very much of the
same order. As an example, for pole--on observations,
the deviations  for $\omega$ = 0.9 are smaller
than 2.5 \%. For an equator--on observer,
 the luminosities which would derived  by the application of the
usual Eq. (1) would be too small by 1.7, 7.5 and 13.7 \%
for $ \omega =$ 0.5, 0.8 and 0.9
respectively. Thus, whatever the weighting function we
consider, we conclude that the errors on the luminosity 
determinations due to rotation effects are very limited.

 \section{Conclusions}

The effects of stellar rotation on the WLR are generally
very  small.
For an average orientation angle, their amplitudes remains 
negligible for most rotation velocities.
 Different orientations angles of the rotation 
axis with respect to the line of sight can produce
some limited scatter on the luminosity
determinations. We estimate the scatter in
the luminosities introduced by orientation effects to be 
at most of the order of 10 \%. 

Here, we have made an analytical approach to the problem.
This does not preclude more complete numerical
simulations as made by Petrenz \& Puls \cite{PP00}), who 
performed 2--D non--LTE models of rotating early--type
stars.

\begin{acknowledgements}
I express my best thanks to Dr. Georges Meynet for
most fruitful discussions.     
\end{acknowledgements}

\end{document}